\documentclass[english]{article}
\usepackage[T1]{fontenc}
\usepackage[latin9]{inputenc}
\usepackage{color}
\usepackage{babel}
\usepackage{latexsym}
\usepackage{refstyle}
\usepackage{float}
\usepackage{url}
\usepackage{amsmath}
\usepackage{amsthm}
\usepackage[unicode=true,pdfusetitle,
 bookmarks=true,bookmarksnumbered=true,bookmarksopen=true,bookmarksopenlevel=1,
 breaklinks=false,pdfborder={0 0 0},pdfborderstyle={},backref=section,colorlinks=true]
 {hyperref}

\makeatletter

\AtBeginDocument{\providecommand\secref[1]{\ref{sec:#1}}}
\let\SF@@footnote\footnote
\def\footnote{\ifx\protect\@typeset@protect
    \expandafter\SF@@footnote
  \else
    \expandafter\SF@gobble@opt
  \fi
}
\expandafter\def\csname SF@gobble@opt \endcsname{\@ifnextchar[
  \SF@gobble@twobracket
  \@gobble
}
\edef\SF@gobble@opt{\noexpand\protect
  \expandafter\noexpand\csname SF@gobble@opt \endcsname}
\def\SF@gobble@twobracket[#1]#2{}
\providecommand{\tabularnewline}{\\}
\floatstyle{ruled}
\newfloat{algorithm}{tbp}{loa}
\providecommand{\algorithmname}{Algorithm}
\floatname{algorithm}{\protect\algorithmname}
\RS@ifundefined{subsecref}
  {\newref{subsec}{name = \RSsectxt}}
  {}
\RS@ifundefined{thmref}
  {\def\RSthmtxt{theorem~}\newref{thm}{name = \RSthmtxt}}
  {}
\RS@ifundefined{lemref}
  {\def\RSlemtxt{lemma~}\newref{lem}{name = \RSlemtxt}}
  {}

\numberwithin{equation}{section}
\numberwithin{figure}{section}
\newcommand{\lyxaddress}[1]{
\par {\raggedright #1
\vspace{1.4em}
\noindent\par}
}

\usepackage{url}

\AtBeginDocument{

}

\makeatother

\usepackage{listings}

\begin{document}

\title{Scaled Rate Optimization for Beta-Binomial Models}

\author{Inon Sharony}
\maketitle

\lyxaddress{\href{mailto:Inon.Sharony@gmail.com}{Inon.Sharony@gmail.com}\footnote{\url{http://sharony.ml}}}

\begin{abstract}
Rates of binomial processes are modeled using beta-binomial distributions
(for example, from Beta Regression). We treat the offline optimization
scenario and then the online one, where we optimize the exploration-exploitation
problem. The rates given by two processes are compared through their
distributions, but we would like to optimize the net payout (given
a constant value per successful event, unique for each of the processes).
The result is an analytically-closed, probabilistic, hypergeometric
expression for comparing the payout distributions of two processes.
To conclude, we contrast this Bayesian result with an alternative
frequentist approach and find 4.5 orders of magnitude improvement
in performance, for a numerical accuracy level of 0.01\%.
\end{abstract}
\begin{description}
\item [{Keywords:}] Bayesian, beta, binomial, hypergeometric, rate
\end{description}

\section{Introduction}

Modeling some proportion quantity is essentially different from independently
modeling a numerator and denominator, and rate prediction is a specific
example of proportion. In some contexts, we would like to compare
two rate processes, which are competing in the context of some portfolio
optimization\cite{41159}. Furthermore, we will assume that our optimization
is performed on time-scales comparable with those of the underlying
rate process, and is therefore treated as an online learning problem.
To treat the offline problem, exact statistical tests can be used.
For binomial processes, the joint probability for the data is given
by the multivariate hypergeometric distribution. See \secref{exact-binomial-data}.

Within the Multi-Armed Bandit picture, a competitor has some intrinsic
payout distribution, and we are tasked with finding an optimal solution
to the exploration-exploitation problem. In this paper, I will derive
an analytically closed expression which optimizes the payout, given
that each competitor has a binomial probability distribution for success,
and some unique payout value for a successful trial \footnote{This value will be assumed to be slowly-changing, relative to all
other time-scales. For example, the rate may fluctuate at an hourly
resolution, or higher, but the payout will change only on the order
of days or weeks.}.

The event of interest is modeled as a binomial process with parameter
$\phi$, where $m$ and $n$ will denote the number of trials and
successes, respectively: $\left(m,n\right)\sim\textnormal{Bin}\left(\phi\right)$.
Therefore, the probability density function (PDF) given the rate parameter
$\phi$, of a potential observation of $n$ wins out of $m$ trials,
is 
\begin{equation}
\Pr\left(m,n|\phi\right)=\left(\begin{array}{c}
m\\
n
\end{array}\right)\phi^{n}\left(1-\phi\right)^{m-n}
\end{equation}

Competing rate processes would manifest in different values for the
rate parameter, $\phi$, which is represented by some unknown underlying
distribution, which we would like to model using empiric data.

\subsection{Beta-binomial model}

To compare rate models, we'd like to compare the probabilities of
the models given some observation data, $\Pr\left(\phi|m,n\right)$.
The conjugate prior of a variable $\left(\phi\right)$ drawn from
a binomial distribution is a beta-binomial distribution, $\phi\sim\textnormal{Beta}\left(\alpha,\beta\right)$:

\begin{equation}
\Pr\left(\phi|\alpha,\beta\right)=\frac{1}{B\left(\alpha,\beta\right)}\phi^{\alpha-1}\left(1-\phi\right)^{\beta-1}
\end{equation}

where $B\left(\cdot\right)$ is the beta function.

The Beta Regression model of Ferrari \& Cribari-Neto\cite{ferrari2004beta}
is used to learn the probabilistic distribution of the rate parameter
$\phi$ of a given binomial process, given its observations. Specifically,
the Beta Regression model is a regression towards the underlying distribution
of $\phi$, from the observed data: $\alpha$ is one plus the observed
number of wins, and $\beta$ is one plus the observed number of losses.
To choose the optimal of two competing rate processes, we compare
the distributions of the rates of these processes.

\subsection{Comparison of rates of beta-binomial processes\cite{evanmiller_bayesian,davidrobinson_betareg}}

Given some observational data $\left(\alpha_{A},\beta_{A},\alpha_{B},\beta_{B}\right)$
for two beta-binomial processes $A$ and $B$, the probability that
the underlying rate of process $B$, $\phi_{B}$, is higher than that
of process $A$, $\phi_{A}$, is (\secref{Prob-comp-rates}):

\begin{align}
\Pr\left(\phi_{B}>\phi_{A}|\alpha_{A},\beta_{A},\alpha_{B},\beta_{B}\right) & =\int_{0}^{1}d\phi_{A}\Pr\left(\phi_{A}|\alpha_{A},\beta_{A}\right)\int_{\phi_{A}}^{1}d\phi_{B}\Pr\left(\phi_{B}|\alpha_{B},\beta_{B}\right)\\
 & =\frac{1}{B\left(\alpha_{A},\beta_{A}\right)}\sum_{i=1}^{\alpha_{B}}\frac{B\left(\alpha_{A}-1+i,\beta_{B}+\beta_{A}\right)}{\left(\beta_{B}-1+i\right)B\left(i,\beta_{B}\right)}
\end{align}

Being combinatoric functions, it is sometimes more convenient to calculate
via the logarithms of the beta functions:
\begin{align}
\Pr\left(\phi_{B}>\phi_{A}|\alpha_{A},\beta_{A},\alpha_{B},\beta_{B}\right) & =\sum_{i=1}^{\alpha_{B}}\exp\left(S\left(\alpha_{A},\beta_{A},\beta_{B},i\right)\right)
\end{align}
\begin{equation}
S\left(\alpha_{A},\beta_{A},\beta_{B},i\right)\equiv\ln B\left(\alpha_{A}-1+i,\beta_{B}+\beta_{A}\right)-\ln B\left(i,\beta_{B}\right)-\ln\left(\beta_{B}-1+i\right)-\ln B\left(\alpha_{A},\beta_{A}\right)
\end{equation}

All terms must enter the exponential (even the prefactor of the sum)
to avoid numerical overflow.

\section{Comparison of payouts of beta-binomial processes}

Since we are interested in optimizing not the success ratio itself,
but the payout (given that one process may have a lower rate, but
higher payout value), we will now derive the expression for $\Pr\left(\phi_{B}>\gamma\phi_{A}|\alpha_{A},\beta_{A},\alpha_{B},\beta_{B}\right)$
given some ratio of the payouts, $\gamma>1$ (without loss of generality).

\begin{equation}
\Pr\left(\phi_{B}>\gamma\phi_{A}|\alpha_{A},\beta_{A},\alpha_{B},\beta_{B}\right)\equiv\int_{0}^{\gamma^{-1}}d\phi_{A}\Pr\left(\phi_{A}|\alpha_{A},\beta_{A}\right)\int_{\gamma\phi_{A}}^{1}d\phi_{B}\Pr\left(\phi_{B}|\alpha_{B},\beta_{B}\right)
\end{equation}
\begin{align}
 & =\int_{0}^{\gamma^{-1}}d\phi_{A}\int_{\gamma\phi_{A}}^{1}d\phi_{B}\frac{\phi_{A}^{\alpha_{A}-1}\left(1-\phi_{A}\right)^{\beta_{A}-1}}{B\left(\alpha_{A},\beta_{A}\right)}\frac{\phi_{B}^{\alpha_{B}-1}\left(1-\phi_{B}\right)^{\beta_{B}-1}}{B\left(\alpha_{B},\beta_{B}\right)}\\
 & =\int_{0}^{\gamma^{-1}}\frac{\phi_{A}^{\alpha_{A}-1}\left(1-\phi_{A}\right)^{\beta_{A}-1}}{B\left(\alpha_{A},\beta_{A}\right)}\int_{\gamma\phi_{A}}^{1}\frac{\phi_{B}^{\alpha_{B}-1}\left(1-\phi_{B}\right)^{\beta_{B}-1}}{B\left(\alpha_{B},\beta_{B}\right)}d\phi_{B}d\phi_{A}\\
 & =\int_{0}^{\gamma^{-1}}\frac{\phi_{A}^{\alpha_{A}-1}\left(1-\phi_{A}\right)^{\beta_{A}-1}}{B\left(\alpha_{A},\beta_{A}\right)}\left[1-I_{\gamma\phi_{A}}\left(\alpha_{B},\beta_{B}\right)\right]d\phi_{A}\\
 & =\int_{0}^{\gamma^{-1}}\frac{\phi_{A}^{\alpha_{A}-1}\left(1-\phi_{A}\right)^{\beta_{A}-1}}{B\left(\alpha_{A},\beta_{A}\right)}\left[1-1+\sum_{i=0}^{\alpha_{B}-1}\frac{\phi_{A}^{i}\left(1-\phi_{A}\right)^{\beta_{B}}}{\left(\beta_{B}+i\right)B\left(1+i,\beta_{B}\right)}\right]d\phi_{A}\\
 & =\sum_{i=0}^{\alpha_{B}-1}\int_{0}^{\gamma^{-1}}\frac{\phi_{A}^{\alpha_{A}-1}\left(1-\phi_{A}\right)^{\beta_{A}-1}}{B\left(\alpha_{A},\beta_{A}\right)}\frac{\phi_{A}^{i}\left(1-\phi_{A}\right)^{\beta_{B}}}{\left(\beta_{B}+i\right)B\left(1+i,\beta_{B}\right)}d\phi_{A}
\end{align}

We now perform a transformation to remove the explicit factor $\gamma$
from the integral boundary: $\gamma\phi_{A}\mapsto\phi_{A^{\prime}}\Rightarrow d\phi_{A}\mapsto\gamma^{-1}d\phi_{A^{\prime}}$.

\begin{align}
\Pr\left(\phi_{B}>\gamma\phi_{A}\right) & =\int_{0}^{1}\gamma^{-1}d\phi_{A^{\prime}}\int_{\phi_{A^{\prime}}}^{1}d\phi_{B}\frac{\gamma^{1-\alpha_{A}}\phi_{A^{\prime}}^{\alpha_{A}-1}\left(1-\phi_{A^{\prime}}/\gamma\right)^{\beta_{A}-1}}{B\left(\alpha_{A},\beta_{A}\right)}\frac{\phi_{B}^{\alpha_{B}-1}\left(1-\phi_{B}\right)^{\beta_{B}-1}}{B\left(\alpha_{B},\beta_{B}\right)}
\end{align}

In appendix \secref{dbl-integral-to-dbl-sum} we show how one of the
integrals, identified as Euler's hypergeometric integral, is solved.

\begin{align}
\Pr\left(\phi_{B}>\gamma\phi_{A}\right) & =\frac{\gamma^{-\alpha_{A}}}{B\left(\alpha_{A},\beta_{A}\right)}\sum_{i=0}^{\alpha_{B}-1}\frac{B\left(\alpha_{A}+i,\beta_{B}+1\right)}{\left(\beta_{B}+i\right)B\left(1+i,\beta_{B}\right)}\ _{2}F_{1}\left(1-\beta_{A},\alpha_{A}+i;\alpha_{A}+i+\beta_{B}+1;\gamma^{-1}\right)
\end{align}

Again, for computational efficiency, we give the logarithmic expression

\begin{align}
\Pr\left(\phi_{B}>\gamma\phi_{A}\right) & =\sum_{i=0}^{\alpha_{B}-1}\exp\left\{ C\left(\alpha_{A},\beta_{A},\gamma\right)+S\left(\alpha_{A}+i,\beta_{B},i\right)+F\left(\alpha_{A}+i,\beta_{A},\beta_{B},\gamma\right)\right\} 
\end{align}
\begin{equation}
C\left(\alpha_{A},\beta_{A},\gamma\right)\equiv-\alpha_{A}\ln\gamma-\ln B\left(\alpha_{A},\beta_{A}\right)
\end{equation}
\begin{align}
S\left(a,\beta_{B},i\right) & \equiv\ln B\left(a,\beta_{B}+1\right)-\ln B\left(1+i,\beta_{B}\right)-\ln\left(\beta_{B}+i\right)
\end{align}

\begin{equation}
F\left(a,\beta_{A},\beta_{B},\gamma\right)=\ln\ _{2}F_{1}\left(1-\beta_{A},a;a+\beta_{B}+1;\gamma^{-1}\right)
\end{equation}

Some computational libraries have direct support of general hypergeometric
functions\footnote{For example, in Python \href{https://docs.scipy.org/doc/scipy/reference/generated/scipy.special.eval_jacobi.html}{SciPy}
and JVM \href{https://mipav.cit.nih.gov/documentation/api/gov/nih/mipav/model/algorithms/Hypergeometric.html}{MIPAV}.}, and other lack it. Luckily for those cases\footnote{e.g. \href{https://commons.apache.org/proper/commons-math/apidocs/org/apache/commons/math4/analysis/polynomials/PolynomialsUtils.html\#createLegendrePolynomial-int-}{Apache Commons Math}.},
our formula is eligible to be implemented using Jacobi polynomials
(\secref{hypergeometric-polynomial})\footnote{In \href{https://docs.scipy.org/doc/scipy/reference/generated/scipy.special.eval_jacobi.html}{SciPy},
the Jacobi polynomials are actually defined in terms of the hypergeometric
function}

\begin{align}
F\left(a,\beta_{A},\beta_{B},\gamma\right) & =\ln P_{\beta_{A}-1}^{\left(a+\beta_{B},\beta_{B}-\beta_{A}+2\right)}\left(1-2\gamma^{-1}\right)+\ln B\left(\beta_{B}-\beta_{A}+1,\beta_{A}-1\right)
\end{align}

It shouldn't be difficult to understand, therefore, how a simple benchmark
of this formula would outperform the equivalent frequentist method
by orders of magnitude.

\section{Results}

A frequentist approach to parameter estimation requires some number
of samples in order to predict the rate with a given level of confidence
(see \footnote{\secref{Sequential-frequentist-AB}}). Since the arrival
time of events is Poisson distributed, the rate at which we can gather
data samples to estimate the rate parameter of each of the competing
processes decays exponentially. 
\begin{quote}
``The advantage of Bayesian formulas over the traditional frequentist
formulas is that you don't have to collect a preordained sample size
in order to get a valid result.''\cite{evanmiller_bayesian}
\end{quote}
Bayesian calculation (see \ref{sec:Numerical-comparison}) shows improvement
of 4.5 orders of magnitude in speed over a Frequentist implementation,
where 10 million samples are required to achieve numerical accuracy
to within 0.01\%, on the random samples generated.

\section{Discussion}

We noted the hypergeometric distribution involved in exactly solving
the offline problem. For the online problem, we derived an analytically-closed,
probabilistic, hypergeometric expression for comparing the payout
distributions of two beta-binomial rate processes. The cost of the
frequentist approach turns out to be prohibitively high for very sparse
data, such as highly-hierarchic or otherwise ``wide'' models.

\section{Acknowledgments}

I would like to thank Bill Tilly for the preliminary exposition, Evan
Miller for posting his analytical formula for comparison of rates
of beta-binomial processes, Chris Stucchio for analyzing its asymptotics,
and to all three for choosing to share their research openly and freely.
I would like to thank Professor Raydonal Ospina Martínez for his encouragement
in writing this report.

\appendix

\section{Exact binomial data\label{sec:exact-binomial-data}}

Given the data in the following contingency table, 

\begin{table}[H]
\begin{tabular}{|c|c|c|c|c|}
\hline 
 & Asset 1 & $\ldots$ & Asset N & Marginal Totals\tabularnewline
\hline 
\hline 
wins & $n_{1}$ & $\ldots$ & $n_{N}$ & $n_{tot.}=\sum_{i}n_{i}$\tabularnewline
\hline 
losses & $o_{1}$ & $\ldots$ & $o_{N}$ & $o_{tot}\equiv\sum_{i}o_{i}$\tabularnewline
\hline 
trials & $m_{1}$ & $\ldots$ & $m_{N}$ & $m_{tot}=\sum_{i}m_{i}$\tabularnewline
\hline 
\end{tabular}

\caption{Contingency table of binomial data}
\end{table}

where $m_{i}=n_{i}+o_{i}$.

\subsection{Fisher's exact test}

The joint probability for the data is given by the multivariate hypergeometric
distribution, $n_{i}\sim HG\left(m_{i},m_{tot},n_{tot}\right)$, and
the exact statistical test is Fisher's exact test.

Denoting the contingency table elements $a_{ij}$ (column-major form),
\begin{equation}
\Pr\left(\left\{ a_{ij}\right\} \right)=\frac{\left(\prod_{i}\left(\left(\sum_{j}a_{ij}\right)!\right)\right)\left(\prod_{j}\left(\left(\sum_{i}a_{ij}\right)!\right)\right)}{\left(\left(\sum_{i,j}a_{ij}\right)!\right)\left(\prod_{i,j}\left(a_{ij}!\right)\right)}
\end{equation}

which in our case reduces to

\begin{equation}
\Pr\left(n_{i};m_{i},m_{tot},n_{tot}\right)=\frac{\left(\begin{array}{c}
n_{tot}\\
n_{i}
\end{array}\right)\left(\begin{array}{c}
m_{tot}-n_{tot}\\
m_{i}-n_{i}
\end{array}\right)}{\left(\begin{array}{c}
m_{tot}\\
m_{i}
\end{array}\right)}
\end{equation}
\begin{align}
 & =\frac{\left(\begin{array}{c}
n_{tot}\\
n_{i}
\end{array}\right)\left(\begin{array}{c}
o_{tot}\\
o_{i}
\end{array}\right)}{\left(\begin{array}{c}
m_{tot}\\
m_{i}
\end{array}\right)}\\
 & =\frac{\left(n_{tot}!o_{tot}!\right)\left(m_{1}!m_{2}!\right)}{\left(m_{tot}!\right)\left(n_{1}!n_{2}!o_{1}!o_{2}!\right)}
\end{align}

\subsection{Significance and confidence}

We would like to compare the data generated by two such models, and
we begin our analysis with the null hypothesis that the two models
have identically distributed underlying rates. The null hypothesis
is rejected if this is supported by observational evidence. That is,
if the probability that the observed evidence combined from both models
(assuming i.i.d. rates) is lower than some significance level (e.g.
p-value lower than 5\%), we can reject the null hypothesis.

\subsubsection{Single-tailed test\protect\footnote{If the evidence provided by the two models is very lopsided, we should
prefer a two-tailed test. For example, if only one of the models is
the incumbent, and therefore we have vastly more observations for
it.}}

If $\sum_{M}\Pr\left(n_{i}^{M};m_{i}^{M},m_{tot}^{M},n_{tot}^{M}\right)<p$,
where $M$ signifies the model, then the null hypothesis can be rejected
on the grounds the evidence provided by the two models differ in a
more extreme way than they would had the models been equivalent.

\subsubsection{Power analysis}

\paragraph{Likelihood-ratio test}

Following the Neyman-Pearson lemma, which states that The most powerful
significance $\left(\alpha\right)$ level test (p-value) is the likelihood
ratio test, we denote the likelihood (and log-likelihood)
\begin{align}
L\left(M;n_{i}^{M},m_{i}^{M},m_{tot}^{M},n_{tot}^{M}\right) & =\Pr\left(n_{i}^{M};m_{i}^{M},m_{tot}^{M},n_{tot}^{M}\right)\\
\ell\left(M;n_{i}^{M},m_{i}^{M},m_{tot}^{M},n_{tot}^{M}\right) & =\ln\Pr\left(n_{i}^{M};m_{i}^{M},m_{tot}^{M},n_{tot}^{M}\right)
\end{align}
\begin{align}
 & =\ln\left(\frac{\Gamma\left(n_{tot}\right)\Gamma\left(o_{tot}\right)}{\Gamma\left(m_{tot}\right)}\frac{\prod_{i}\Gamma\left(m_{i}\right)}{\prod_{i}\Gamma\left(n_{i}\right)\prod_{i}\Gamma\left(o_{i}\right)}\right)\\
 & =\ln\left(\frac{\textnormal{Beta}\left(n_{tot},o_{tot}\right)}{\prod_{i}\textnormal{Beta}\left(n_{i},o_{i}\right)}\right)\\
 & =\ln\textnormal{Beta}\left(n_{tot},o_{tot}\right)-\sum_{i}\ln\textnormal{Beta}\left(n_{i},o_{i}\right)
\end{align}

\subparagraph{Comparing models using Wilk's theorem}

Define the alternative hypothesis as the model with more degrees of
freedom, $\nu_{D}\equiv\nu_{1}-\nu_{0}\ge0$, and the test statistic
$D$

\begin{equation}
D=-2\ln\Lambda=2\left[\ell\left(M=H_{1}\right)-\ell\left(M=H_{0}\right)\right]
\end{equation}

The probability distribution of $D$ tends to a $\chi_{\nu_{D}}^{2}$
distribution as the sample size tends to infinity.

The use of this theorem is in approximating the limit of the p-value
for large sample sets, via the tabulated probability density distribution
of $\lim_{m_{tot}\rightarrow\infty}\Pr\left(D\right)=\chi_{\nu_{D}}^{2}\left(D\right)$.

Note that the test statistic here is chi-squared, which makes some
assumptions on the distribution of the samples.

\paragraph{A non-parametric test: Kolmogorov-Smirnov}

For a compared quantity $x$, we denote the empirical CDF (ECDF) of
$x$ (for example from a histogram of $x$) by $F_{K}\left(x\right)$,
where $K\in\left\{ A,B\right\} $ and $m_{K}$ is the number of impressions
given to $K$. The Kolmogorov-Smirnov statistic is

\begin{equation}
D_{m_{A},m_{B}}\equiv\sup_{x}\left|F_{B}\left(x\right)-F_{A}\left(x\right)\right|
\end{equation}

The null hypothesis (B is not different from A) is rejected at level
$\alpha$ if
\begin{align}
D_{m_{A},m_{B}} & >c\left(\alpha\right)\sqrt{\frac{\sum_{K}m_{K}}{\prod_{K}m_{K}}}\\
c\left(\alpha\right) & \equiv\sqrt{-\frac{1}{2}\ln\left(\frac{\alpha}{2}\right)}
\end{align}

\section{Probabilistic comparison of the rates of two processes\label{sec:Prob-comp-rates}}

The probability, given some observational data $\left(\alpha_{A},\beta_{A},\alpha_{B},\beta_{B}\right)$
for two beta-binomial processes $A$ and $B$, that the underlying
rate of process $B$, $\phi_{B}$, is higher than that of process
$A$, $\phi_{A}$, is:

\begin{equation}
\Pr\left(\phi_{B}>\phi_{A}|\alpha_{A},\beta_{A},\alpha_{B},\beta_{B}\right)=\int_{0}^{1}d\phi_{A}\Pr\left(\phi_{A}|\alpha_{A},\beta_{A}\right)\int_{\phi_{A}}^{1}d\phi_{B}\Pr\left(\phi_{B}|\alpha_{B},\beta_{B}\right)
\end{equation}
\begin{align}
 & =\int_{0}^{1}d\phi_{A}\int_{\phi_{A}}^{1}d\phi_{B}\frac{\phi_{A}^{\alpha_{A}-1}\left(1-\phi_{A}\right)^{\beta_{A}-1}}{B\left(\alpha_{A},\beta_{A}\right)}\frac{\phi_{B}^{\alpha_{B}-1}\left(1-\phi_{B}\right)^{\beta_{B}-1}}{B\left(\alpha_{B},\beta_{B}\right)}\\
 & =\int_{0}^{1}d\phi_{A}\left[\frac{\phi_{A}^{\alpha_{A}-1}\left(1-\phi_{A}\right)^{\beta_{A}-1}}{B\left(\alpha_{A},\beta_{A}\right)}\int_{\phi_{A}}^{1}d\phi_{B}\frac{\phi_{B}^{\alpha_{B}-1}\left(1-\phi_{B}\right)^{\beta_{B}-1}}{B\left(\alpha_{B},\beta_{B}\right)}\right]\\
 & =1-\int_{0}^{1}\frac{\phi_{A}^{\alpha_{A}-1}\left(1-\phi_{A}\right)^{\beta_{A}-1}}{B\left(\alpha_{A},\beta_{A}\right)}I_{\phi_{A}}\left(\alpha_{B},\beta_{B}\right)d\phi_{A}
\end{align}

where $I_{\phi}\left(\alpha,\beta\right)$ is just shorthand for the
regularized incomplete beta function. We next use a lemma (\secref{Lemma-reg-incompl-beta-fun})
to simplify this expression to
\begin{align}
\Pr\left(\phi_{B}>\phi_{A}\right) & =1-\int_{0}^{1}\frac{\phi_{A}^{\alpha_{A}-1}\left(1-\phi_{A}\right)^{\beta_{A}-1}}{B\left(\alpha_{A},\beta_{A}\right)}\left(1-\sum_{i=0}^{\alpha_{B}-1}\frac{\phi_{A}^{i}\left(1-\phi_{A}\right)^{\beta_{B}}}{\left(\beta_{B}+i\right)B\left(i+1,\beta_{B}\right)}\right)d\phi_{A}
\end{align}
\begin{align}
 & =1-1+\int_{0}^{1}\frac{\phi_{A}^{\alpha_{A}-1}\left(1-\phi_{A}\right)^{\beta_{A}-1}}{B\left(\alpha_{A},\beta_{A}\right)}\sum_{i=0}^{\alpha_{B}-1}\frac{\phi_{A}^{i}\left(1-\phi_{A}\right)^{\beta_{B}}}{\left(\beta_{B}+i\right)B\left(i+1,\beta_{B}\right)}d\phi_{A}\\
 & =\sum_{i=0}^{\alpha_{B}-1}\int_{0}^{1}\frac{\phi_{A}^{\alpha_{A}+i-1}\left(1-\phi_{A}\right)^{\beta_{A}+\beta_{B}-1}}{\left(\beta_{B}+i\right)B\left(\alpha_{A},\beta_{A}\right)B\left(i+1,\beta_{B}\right)}d\phi_{A}\\
 & =\sum_{i=0}^{\alpha_{B}-1}\frac{1}{\left(\beta_{B}+i\right)B\left(\alpha_{A},\beta_{A}\right)B\left(i+1,\beta_{B}\right)}\int_{0}^{1}\phi_{A}^{\alpha_{A}+i-1}\left(1-\phi_{A}\right)^{\beta_{A}+\beta_{B}-1}d\phi_{A}\\
 & =\sum_{i=0}^{\alpha_{B}-1}\frac{B\left(\alpha_{A}+i,\beta_{A}+\beta_{B}\right)}{\left(\beta_{B}+i\right)B\left(\alpha_{A},\beta_{A}\right)B\left(i+1,\beta_{B}\right)}\int_{0}^{1}\frac{\phi_{A}^{\alpha_{A}+i-1}\left(1-\phi_{A}\right)^{\beta_{A}+\beta_{B}-1}}{B\left(\alpha_{A}+i,\beta_{A}+\beta_{B}\right)}d\phi_{A}
\end{align}

where in the last line we multiplied and divided by $B\left(\alpha_{A}+i,\beta_{A}+\beta_{B}\right)$,
and then the integral term is just the integral of the distribution
\textbf{$\textnormal{Beta}\left(\alpha_{A}+i,\beta_{A}+\beta_{B}\right)$},
which is unity.

Re-indexing,

\begin{align}
\Pr\left(\phi_{B}>\phi_{A}|\alpha_{A},\beta_{A},\alpha_{B},\beta_{B}\right) & =\sum_{i=0}^{\alpha_{B}-1}\frac{B\left(\alpha_{A}+i,\beta_{B}+\beta_{A}\right)}{\left(\beta_{B}+i\right)B\left(1+i,\beta_{B}\right)B\left(\alpha_{A},\beta_{A}\right)}
\end{align}
\begin{align}
 & =\frac{1}{B\left(\alpha_{A},\beta_{A}\right)}\sum_{i=1}^{\alpha_{B}}\frac{B\left(\alpha_{A}-1+i,\beta_{B}+\beta_{A}\right)}{\left(\beta_{B}-1+i\right)B\left(i,\beta_{B}\right)}
\end{align}

Chris Stucchio has published an asymptotic analysis of this formula\cite{stucchio_asymptotic}.

\section{Lemma for the regularized incomplete beta function\label{sec:Lemma-reg-incompl-beta-fun}}

Recursively iterating

\begin{equation}
I_{x}\left(\alpha,\beta\right)=I_{x}\left(\alpha-1,\beta\right)-\frac{x^{\alpha-1}\left(1-x\right)^{\beta}}{\left(\alpha-1\right)B\left(\alpha-1,\beta\right)}
\end{equation}

Until the base case

\begin{equation}
I_{x}\left(1,\beta\right)=1-\left(1-x\right)^{\beta}
\end{equation}

We get

\begin{align}
I_{x}\left(\alpha,\beta\right) & =1-\left(1-x\right)^{\beta}-\sum_{i=1}^{\alpha-1}\frac{x^{\alpha-i}\left(1-x\right)^{\beta}}{\left(\alpha-i\right)B\left(\alpha-i,\beta\right)}
\end{align}

Subsuming the zeroth term into the sum:

\begin{equation}
I_{x}\left(\alpha,\beta\right)=1-\sum_{i=0}^{\alpha-1}\frac{x^{i}\left(1-x\right)^{\beta}}{\left(\beta+i\right)B\left(1+i,\beta\right)}
\end{equation}

\section{Solution of Euler's hypergeometric integral\label{sec:dbl-integral-to-dbl-sum}}

Starting from

\begin{align}
\Pr\left(\phi_{B}>\gamma\phi_{A}\right) & =\gamma^{-\alpha_{A}}\int_{0}^{1}d\phi_{A^{\prime}}\int_{\phi_{A^{\prime}}}^{1}d\phi_{B}\frac{\phi_{A^{\prime}}^{\alpha_{A}-1}\left(1-\phi_{A^{\prime}}/\gamma\right)^{\beta_{A}-1}}{B\left(\alpha_{A},\beta_{A}\right)}\frac{\phi_{B}^{\alpha_{B}-1}\left(1-\phi_{B}\right)^{\beta_{B}-1}}{B\left(\alpha_{B},\beta_{B}\right)}
\end{align}

we now employ the (exact) binomial expansion:

\begin{equation}
\left(1-\phi_{A^{\prime}}/\gamma\right)^{\beta_{A}-1}=\sum_{k=0}^{\beta_{A}-1}\left(\begin{array}{c}
\beta_{A}-1\\
k
\end{array}\right)\left(-\phi_{A^{\prime}}/\gamma\right)^{k}=\sum_{k=0}^{\beta_{A}-1}\left(\begin{array}{c}
\beta_{A}-1\\
k
\end{array}\right)\left(-1\right)^{k}\phi_{A^{\prime}}^{k}\gamma^{-k}
\end{equation}

so:

\begin{align}
 & \Pr\left(\phi_{B}>\gamma\phi_{A}\right)=\\
 & =\gamma^{-\alpha_{A}}\int_{0}^{1}d\phi_{A^{\prime}}\int_{p_{A^{\prime}}}^{1}d\phi_{B}\frac{\phi_{A^{\prime}}^{\alpha_{A}-1}}{B\left(\alpha_{A},\beta_{A}\right)}\sum_{k=0}^{\beta_{A}-1}\left(-1\right)^{k}\left(\begin{array}{c}
\beta_{A}-1\\
k
\end{array}\right)\phi_{A^{\prime}}^{k}\gamma^{-k}\frac{\phi_{B}^{\alpha_{B}-1}\left(1-\phi_{B}\right)^{\beta_{B}-1}}{B\left(\alpha_{B},\beta_{B}\right)}\\
 & =\gamma^{-\alpha_{A}}\int_{0}^{1}d\phi_{A^{\prime}}\int_{p_{A^{\prime}}}^{1}d\phi_{B}\sum_{k=0}^{\beta_{A}-1}\left(-1\right)^{k}\left(\begin{array}{c}
\beta_{A}-1\\
k
\end{array}\right)\gamma^{-k}\frac{\phi_{A^{\prime}}^{\alpha_{A}+k-1}}{B\left(\alpha_{A},\beta_{A}\right)}\frac{\phi_{B}^{\alpha_{B}-1}\left(1-\phi_{B}\right)^{\beta_{B}-1}}{B\left(\alpha_{B},\beta_{B}\right)}\\
 & =\sum_{k=0}^{\beta_{A}-1}\left(-1\right)^{k}\left(\begin{array}{c}
\beta_{A}-1\\
k
\end{array}\right)\frac{\gamma^{-k-\alpha_{A}}}{B\left(\alpha_{A},\beta_{A}\right)}\int_{0}^{1}d\phi_{A^{\prime}}\phi_{A^{\prime}}^{\alpha_{A}+k-1}\left[1-I_{\phi_{A^{\prime}}}\left(\alpha_{B},\beta_{B}\right)\right]
\end{align}

using the same lemma from \secref{Lemma-reg-incompl-beta-fun},

\begin{equation}
I_{\phi_{A^{\prime}}}\left(\alpha_{B},\beta_{B}\right)=1-\sum_{i=0}^{\alpha_{B}-1}\frac{\phi_{A^{\prime}}^{i}\left(1-\phi_{A^{\prime}}\right)^{\beta_{B}}}{\left(\beta_{B}+i\right)B\left(1+i,\beta_{B}\right)}
\end{equation}

we get

\begin{align}
 & \Pr\left(\phi_{B}>\gamma\phi_{A}\right)=\\
 & =\sum_{k=0}^{\beta_{A}-1}\left(-1\right)^{k}\left(\begin{array}{c}
\beta_{A}-1\\
k
\end{array}\right)\frac{\gamma^{-k-\alpha_{A}}}{B\left(\alpha_{A},\beta_{A}\right)}\int_{0}^{1}d\phi_{A^{\prime}}\phi_{A^{\prime}}^{\alpha_{A}+k-1}\left[1-1+\sum_{i=0}^{\alpha_{B}-1}\frac{\phi_{A^{\prime}}^{i}\left(1-\phi_{A^{\prime}}\right)^{\beta_{B}}}{\left(\beta_{B}+i\right)B\left(1+i,\beta_{B}\right)}\right]\\
 & =\sum_{k=0}^{\beta_{A}-1}\sum_{i=0}^{\alpha_{B}-1}\left(-1\right)^{k}\left(\begin{array}{c}
\beta_{A}-1\\
k
\end{array}\right)\frac{\gamma^{-k-\alpha_{A}}}{B\left(\alpha_{A},\beta_{A}\right)}\int_{0}^{1}d\phi_{A^{\prime}}\frac{\phi_{A^{\prime}}^{\alpha_{A}+k+i-1}\left(1-\phi_{A^{\prime}}\right)^{\beta_{B}}}{\left(\beta_{B}+i\right)B\left(1+i,\beta_{B}\right)}
\end{align}

we define $\alpha_{A^{\prime}}\equiv\alpha_{A}+k+i$ and $\beta_{B^{\prime}}\equiv\beta_{B}+1$

\begin{align}
 & \Pr\left(\phi_{B}>\gamma\phi_{A}\right)=\\
 & =\sum_{k=0}^{\beta_{A}-1}\sum_{i=0}^{\alpha_{B}-1}\left(-1\right)^{k}\left(\begin{array}{c}
\beta_{A}-1\\
k
\end{array}\right)\frac{\gamma^{-k-\alpha_{A}}}{\left(\beta_{B}+i\right)B\left(1+i,\beta_{B}\right)B\left(\alpha_{A},\beta_{A}\right)}\int_{0}^{1}d\phi_{A^{\prime}}\phi_{A^{\prime}}^{\alpha_{A^{\prime}}-1}\left(1-\phi_{A^{\prime}}\right)^{\beta_{B^{\prime}}-1}\\
 & =\sum_{k=0}^{\beta_{A}-1}\sum_{i=0}^{\alpha_{B}-1}\left(-1\right)^{k}\left(\begin{array}{c}
\beta_{A}-1\\
k
\end{array}\right)\frac{\gamma^{-k-\alpha_{A}}B\left(\alpha_{A^{\prime}},\beta_{B^{\prime}}\right)}{\left(\beta_{B}+i\right)B\left(1+i,\beta_{B}\right)B\left(\alpha_{A},\beta_{A}\right)}
\end{align}

where in the last step we multiplied and divided by $B\left(\alpha_{A^{\prime}},\beta_{B^{\prime}}\right)$
and integrated over the entire $\textnormal{Beta\ensuremath{\left(\alpha_{A^{\prime}},\beta_{B^{\prime}}\right)}}$
distribution to get unity.

\begin{align}
 & \Pr\left(\phi_{B}>\gamma\phi_{A}\right)=\\
 & =\frac{\gamma^{-\alpha_{A}}}{B\left(\alpha_{A},\beta_{A}\right)}\sum_{i=0}^{\alpha_{B}-1}\frac{1}{\left(\beta_{B}+i\right)B\left(1+i,\beta_{B}\right)}\sum_{k=0}^{\beta_{A}-1}\left(-\gamma\right)^{-k}\left(\begin{array}{c}
\beta_{A}-1\\
k
\end{array}\right)B\left(\alpha_{A}+i+k,\beta_{B}+1\right)
\end{align}

Using a definition of the Gauss hypergeometric series, detailed in
the appendix \secref{Hypergeometric-series}, we identify the second
sum as proportional to the Gauss hypergeometric function

\begin{align}
\Pr\left(\phi_{B}>\gamma\phi_{A}\right) & =\frac{\gamma^{-\alpha_{A}}}{B\left(\alpha_{A},\beta_{A}\right)}\sum_{i=0}^{\alpha_{B}-1}\frac{B\left(\alpha_{A}+i,\beta_{B}+1\right)}{\left(\beta_{B}+i\right)B\left(1+i,\beta_{B}\right)}\ _{2}F_{1}\left(1-\beta_{A},\alpha_{A}+i;\alpha_{A}+i+\beta_{B}+1;\gamma^{-1}\right)
\end{align}

\section{Hypergeometric series\label{sec:Hypergeometric-series}}

The hypergeometric function is defined for $\left|z\right|<1$ by
the power series

\begin{equation}
_{2}F_{1}\left(x_{1},x_{2};y;z\right)=\sum_{n=0}^{\infty}\frac{\left(x_{1}\right)_{n}^{+}\left(x_{2}\right)_{n}^{+}}{\left(y\right)_{n}^{+}}\frac{z^{n}}{n!}
\end{equation}

where $\left(q\right)_{n}^{+}$ is the rising factorial or Pochhammer
symbol (written to avoid confusion with $\left(q\right)_{n}$ which
also refers to the falling factorial)
\begin{align}
\left(q\right)_{n} & \equiv\frac{\Gamma\left(q+n\right)}{\Gamma\left(q\right)}
\end{align}

Using the following series expansion for a hypergeometric function
with a non-positive integer parameter:
\begin{align}
\sum_{n=0}^{m}\left(-1\right)^{n}\left(\begin{array}{c}
m\\
n
\end{array}\right)\frac{\left(x_{2}\right)_{n}^{+}}{\left(y\right)_{n}^{+}}z^{n} & =_{2}F_{1}\left(-m,x_{2};y;z\right)
\end{align}

in our case, $\left(m=\beta_{A}-1,x_{2}=\alpha_{A}+i;y=\alpha_{A}+i+\beta_{B}+1;z=\gamma^{-1}\right)$,
and we will introduce the following variables for convenience: $a\equiv\alpha_{A}+i$
and $b\equiv\beta_{B}+1$.

\begin{align}
_{2}F_{1}\left(-\left(\beta_{A}-1\right),a;a+b;\gamma^{-1}\right) & =\sum_{k=0}^{\beta_{A}-1}\left(-1\right)^{k}\left(\begin{array}{c}
\beta_{A}-1\\
k
\end{array}\right)\frac{\left(a\right)_{k}^{+}}{\left(a+b\right)_{k}^{+}}\gamma^{-k}
\end{align}
\begin{align}
 & =\sum_{k=0}^{\beta_{A}-1}\left(-\gamma\right)^{-k}\left(\begin{array}{c}
\beta_{A}-1\\
k
\end{array}\right)\frac{\Gamma\left(a+k\right)}{\Gamma\left(a\right)}\frac{\Gamma\left(a+b\right)}{\Gamma\left(a+b+k\right)}\\
 & =\frac{\Gamma\left(a+b\right)}{\Gamma\left(a\right)}\sum_{k=0}^{\beta_{A}-1}\left(-\gamma\right)^{-k}\left(\begin{array}{c}
\beta_{A}-1\\
k
\end{array}\right)\frac{\Gamma\left(a+k\right)}{\Gamma\left(a+b+k\right)}\\
 & =\frac{\Gamma\left(b\right)}{B\left(a,b\right)}\sum_{k=0}^{\beta_{A}-1}\left(-\gamma\right)^{-k}\left(\begin{array}{c}
\beta_{A}-1\\
k
\end{array}\right)\frac{B\left(a+k,b\right)}{\Gamma\left(b\right)}\\
 & =\frac{1}{B\left(a,b\right)}\sum_{k=0}^{\beta_{A}-1}\left(-\gamma\right)^{-k}\left(\begin{array}{c}
\beta_{A}-1\\
k
\end{array}\right)B\left(a+k,b\right)
\end{align}

\section{Evaluation of the hypergeometric function using a Jacobi polynomial\label{sec:hypergeometric-polynomial}}

We start from the following identity

\begin{align}
\ _{2}F_{1}\left(-m,m+x+1+y;x+1;z\right) & =\frac{m!}{\left(x+1\right)_{m}^{+}}P_{m}^{\left(x,y\right)}\left(1-2z\right)
\end{align}

Next, we use the following variable transformations to recover the
hypergeometric function in the form we used above
\begin{align*}
a & =-m\Rightarrow m=-a\\
c & =y+1\Rightarrow y=c-1\\
b & =m+x+1+y\Rightarrow x=b-m-y-1=b+a-c+1-1=b+a-c
\end{align*}

and now we have

\begin{align}
\ _{2}F_{1}\left(a,b;c;z\right) & =\frac{\left(-a\right)!}{\left(c\right)_{-a}^{+}}P_{-a}^{\left(c-1,b+a-c\right)}\left(1-2z\right)
\end{align}
\begin{align}
 & =\frac{\Gamma\left(-a\right)}{\Gamma\left(c-a\right)/\Gamma\left(c\right)}P_{-a}^{\left(c-1,b+a-c\right)}\left(1-2z\right)\\
 & =B\left(c,-a\right)P_{-a}^{\left(c-1,b+a-c\right)}\left(1-2z\right)
\end{align}

for $\ _{2}F_{1}\left(1-\beta_{A},a;a+\beta_{B}+1;\gamma^{-1}\right)$
we find the following variable identities:

\begin{align*}
z & =\gamma^{-1}\\
m & =\beta_{A}-1\\
x+1 & =a+\beta_{B}+1\Rightarrow x=a+\beta_{B}\\
m+x+1+y & =a\Rightarrow y=a-m-x-1=a-\beta_{A}+1-a-\beta_{B}+1=\beta_{B}-\beta_{A}+2
\end{align*}

And finally we have

\begin{align}
\ _{2}F_{1}\left(1-\beta_{A},a;a+\beta_{B}+1;\gamma^{-1}\right) & =\frac{\left(\beta_{A}-1\right)!}{\left(\beta_{B}-\beta_{A}+1\right)_{\beta_{A}-1}^{+}}P_{\beta_{A}-1}^{\left(a+\beta_{B},\beta_{B}-\beta_{A}+2\right)}\left(1-2\gamma^{-1}\right)
\end{align}
\begin{align}
 & =\Gamma\left(\beta_{A}-1\right)\frac{\Gamma\left(\beta_{B}-\beta_{A}+1\right)}{\Gamma\left(\beta_{B}-\beta_{A}+1+\beta_{A}-1\right)}P_{\beta_{A}-1}^{\left(a+\beta_{B},\beta_{B}-\beta_{A}+2\right)}\left(1-2\gamma^{-1}\right)\\
 & =B\left(\beta_{B}-\beta_{A}+1,\beta_{A}-1\right)P_{\beta_{A}-1}^{\left(a+\beta_{B},\beta_{B}-\beta_{A}+2\right)}\left(1-2\gamma^{-1}\right)
\end{align}

\section{Sequential frequentist approach\cite{evanmiller_frequentist}\label{sec:Sequential-frequentist-AB}}
\begin{quote}
The key insight in Ben Tilly's article\cite{tilly_multiplelooks}
is that if users are randomly assigned to two groups, and the two
groups have the same conversion rate, then the sequence of successes
from the two groups is mathematically equivalent to a series of random
coin flips.
\end{quote}
The following procedure is derived from the analysis of the gambler's
ruin problem for this one-dimensional random walk,

\begin{algorithm}[H]
\begin{enumerate}
\item At the beginning of the experiment, choose a sample size $N$.
\item Assign subjects randomly to the treatment and control, with 50\% probability
each.
\item Track the number of incoming successes from the treatment group. Call
this number $T\equiv n_{tot}^{M}$.
\item Track the number of incoming successes from the control group. Call
this number $C\equiv n_{tot}-n_{tot}^{M}$.
\item If $d\equiv T-C=n_{tot}^{M}-\left(n_{tot}-n_{tot}^{M}\right)=2n_{tot}^{M}-n_{tot}$
reaches $2\sqrt{N}$, stop the test. Declare the treatment to be the
winner.
\item If $n_{tot}=T+C$ reaches $N$, stop the test. Declare no winner.
\end{enumerate}
\caption{Simple sequential A/B testing (Evan Miller)}
\end{algorithm}

A reference to the proof for step 5 is given in the original post.

Samples should be i.i.d. between the models (C and T). Ideally, online
testing of T vs C should be done on mutually exclusive sets to avoid
effects of interactions between C and T.

\subsection{Power and significance}

Given a model with $n_{tot}^{M}$ total wins such that $n_{tot}=n_{tot}^{M}+\left(n_{tot}^{M}-d_{M}^{*}\right)$
and that the sum of the wins by both models is $n_{tot}$.

\begin{align}
\alpha & >\sum_{j=1}^{n_{tot}}\frac{n_{tot}^{M}}{j}\left(\begin{array}{c}
j\\
\left(d_{M}^{*}+j\right)/2
\end{array}\right)2^{-j}\\
\beta & >1-\sum_{j=1}^{n_{tot}}\frac{n_{tot}^{M}}{j}\left(\begin{array}{c}
j\\
\left(d_{M}^{*}+j\right)/2
\end{array}\right)\left(\frac{1}{2+\delta_{M}}\right)^{\left(j-d_{M}^{*}\right)/2}\left(\frac{1+\delta_{M}}{2+\delta_{M}}\right)^{\left(j+d_{M}^{*}\right)/2}\\
 & =1-\sum_{j=1}^{n_{tot}}\frac{n_{tot}^{M}}{j}\left(\begin{array}{c}
j\\
\left(d_{M}^{*}+j\right)/2
\end{array}\right)\left(2+\delta_{M}\right)^{-j}\left(1+\delta_{M}\right)^{\left(d_{M}^{*}+j\right)/2}
\end{align}

where $\delta_{M}=n_{tot}^{M}/\left(n_{tot}-n_{tot}^{M}\right)$ is
the lift.

For example, for $\alpha=5\%$, $\beta=20\%$ and $\delta=50\%$,
we get $n_{tot}=170$ and $d_{M}^{*}=26$.

So, to see if we can get 50\% lift with a p-value of 5\% and 80\%
power, we should look for a 26 win margin in favor of the treatment,
or give up if we reach 170 overall wins.

\section{Numerical comparison of Bayesian vs. frequentist calculations\label{sec:Numerical-comparison}}

\begin{lstlisting}
from time import perf_counter

from numpy import exp, log, mean, nan, reciprocal
from numpy.random import beta, random
from scipy.special import betaln, binom, hyp2f1


def frequentist(alpha_a, beta_a, alpha_b, beta_b, gamma, n):
    return mean(beta(alpha_b, beta_b, size=n) > gamma * beta(alpha_a, beta_a, size=n))


def pr_b_gt_pr_ga(alpha_a, beta_a, alpha_b, beta_b, gamma):
    assert gamma > 1
    result = 0
    m = beta_a - 1
    b = beta_b + 1
    z = 1 / gamma
    c = -alpha_a * log(gamma) - betaln(alpha_a, beta_a)
    for i in range(alpha_b):
        a = alpha_a + i
        s = betaln(a, b) - betaln(i + 1, beta_b) - log(beta_b + i)
        f = log(hyp2f1(-m, a, a + b, z))
        result += exp(c + s + f)
    return result


def main():
    i, j = 0, 0
    hg_times, freq_times = 0, 0
    while j < 10:
        alpha_a, beta_a, alpha_b, beta_b, gamma = map(lambda x: int(x + 1), reciprocal(random(5)))
        i += 1
        if alpha_a > beta_a or alpha_b > beta_b:
            continue
        start = perf_counter()
        hg = pr_b_gt_pr_ga(alpha_a, beta_a, alpha_b, beta_b, gamma)
        stop = perf_counter()
        hg_times += stop - start
        if hg is None or hg == nan:
            continue
        print("hg: ", hg, stop - start)
        start = perf_counter()
        freq = frequentist(alpha_a, beta_a, alpha_b, beta_b, gamma, pow(10, 7))
        stop = perf_counter()
        freq_times += stop - start
        print("freq ", freq, stop - start)
        j += 1
    print(hg_times / j, freq_times / j)
\end{lstlisting}

\bibliographystyle{unsrt}
\phantomsection\addcontentsline{toc}{section}{\refname}\bibliography{abtesting}

\begin{thebibliography}{1}

\bibitem{41159}
H.~Brendan McMahan, Gary Holt, D.~Sculley, Michael Young, Dietmar Ebner, Julian
  Grady, Lan Nie, Todd Phillips, Eugene Davydov, Daniel Golovin, Sharat
  Chikkerur, Dan Liu, Martin Wattenberg, Arnar~Mar Hrafnkelsson, Tom Boulos,
  and Jeremy Kubica.
\newblock Ad click prediction: a view from the trenches.
\newblock In {\em Proceedings of the 19th ACM SIGKDD International Conference
  on Knowledge Discovery and Data Mining (KDD)}, 2013.

\bibitem{ferrari2004beta}
Silvia Ferrari and Francisco Cribari-Neto.
\newblock Beta regression for modelling rates and proportions.
\newblock {\em Journal of Applied Statistics}, 31(7):799--815, 2004.
\newblock Taylor \& Francis.

\bibitem{evanmiller_bayesian}
Evan Miller.
\newblock Formulas for bayesian a/b testing.
\newblock "\url{http://www.evanmiller.org/bayesian-ab-testing.html}", October
  2015.

\bibitem{davidrobinson_betareg}
David Robinson.
\newblock Understanding beta binomial regression (using baseball statistics).
\newblock "\url{http://varianceexplained.org/r/beta_binomial_baseball}", May
  2016.

\bibitem{stucchio_asymptotic}
Chris Stucchio.
\newblock Asymptotics of evan miller's bayesian a/b formula.
\newblock
  "\url{https://www.chrisstucchio.com/blog/2014/bayesian_asymptotics.html}",
  2014.

\bibitem{evanmiller_frequentist}
Evan Miller.
\newblock Simple sequential a/b testing.
\newblock "\url{http://www.evanmiller.org/sequential-ab-testing.html}", October
  2015.

\bibitem{tilly_multiplelooks}
Ben Tilly.
\newblock A/b testing with multiple looks.
\newblock "\url{http://elem.com/~btilly/ab-testing-multiple-looks/}".

\end{thebibliography}

\end{document}